# A multi-frequency high-field pulsed EPR / ENDOR spectrometer

Gavin W. Morley, Louis-Claude Brunel, and Johan van Tol
National High Magnetic Field Laboratory at Florida State University, Tallahassee, Florida 32310, USA

We describe a pulsed multi-frequency electron paramagnetic resonance spectrometer operating at several frequencies in the range of 110-336 GHz. The microwave source at all frequencies consists of a multiplier chain starting from a solid state synthesizer in the 12-15 GHz range. A fast PIN-switch at the base frequency creates the pulses. At all frequencies a Fabry-Pérot resonator is employed and the π/2 pulse length ranges from ~100 ns at 110 GHz to ~600 ns at 334 GHz. Measurements of a single crystal containing dilute $Mn^{2+}$ impurities at 12 T illustrate the effects of large electron spin polarizations. The capabilities also allow for pulsed electron nuclear double resonance experiments as demonstrated by Mims ENDOR of $^{39}K$ nuclei in $Cr:K_3NbO_8$.

## I. INTRODUCTION

Electron paramagnetic resonance (EPR) experiments have realized new opportunities as magnetic fields have been increased[1,2,3,4,5,6,7] and pulsed capabilities have been added[8,9,10,11,12,13,14].

Pulsed magnetic resonance probes spin dynamics directly in a way that continuous-wave (cw) measurements cannot[15]. For example, spin decoherence times ($T_2$) are a standard result of pulsed experiments. The recent interest in quantum information has stimulated research into pulsed EPR, as it can be used for qubit gate operations as well as qubit initialization and measurement[16,17,18,19,20]. However, when $T_2$ times are short compared with the time needed to flip a spin, cw EPR provides more information than pulsed experiments. The spectrometer described here can be changed between cw and pulsed modes in less than five minutes without removing the sample.

High field/high frequency EPR has the following advantages: (i) At higher fields an EPR experiment can resolve more resonances, such as those from multiple spin centres and anisotropic g-tensors[1,14]. (ii) Resonances that occur at high frequencies in zero magnetic field (as found in samples with strong spin-orbit coupling) can only be studied with the high frequencies that are used in a high-field spectrometer[6]. (iii) The electronic polarization is enhanced: for an electron spin with $g = 2$ at 12 T, the Zeeman energy is much greater than the thermal energy at temperatures as high as 4.2 K. This electronic polarization has been used to polarize nuclei[18,19]. (iv) Higher-field spectrometers have larger absolute spin sensitivity[9] as long as the microwave power is high enough to excite all of the spins. Resonators tend to get smaller as the frequency is increased, so may be better suited to small samples. (v) Electron-nuclear double resonance (ENDOR) spectra benefit from more spectral resolution at high fields[21], as is the case for NMR. (vi) The ENDOR sensitivity to nuclei with low gyromagnetic ratios is higher when the ENDOR frequency is raised by increasing the magnetic field[15].

The microwave sources used for high frequency EPR include orotrons[12], gyrotrons[2], extended interaction klystrons[11], far infra-red lasers[3,10], Gunn diodes[1,13,14] and IMPATT phase locked oscillators[9]. The frequency and power available from amplified synthesizers with multipliers have increased in recent years, and such sources were used in the spectrometer described here, supplied by Virginia Diodes Inc. (VDI)[22].

## II. EXPERIMENTAL

The quasi-optic[23] induction mode[6] design of our cw spectrometer has been described previously[1]. It employs a Martin-Pupplett interferometer which can vary the polarization of the millimeter waves incident on the sample from linear to circular polarization. In cw-mode, where available power is not an issue, the highest sensitivity can be obtained by applying linearly polarized microwaves and detecting the signal in the cross-polarized reflected component.

For pulsed EPR the oscillating magnetic field, $B_1$, must be large enough to flip the electron spins before they dephase. In the spectrometer described here the millimeter waves travel along the direction of the static magnetic field, so the largest useful $B_1$ field is obtained with the correct circular polarization. However, the reflected power of the pulses then has the same polarization as the response from the sample, and cannot be easily separated by passive components. The intensity of the reflected pulses is above the damage limit of the mixer detector, and therefore in general linearly polarized millimeter wave pulses are employed, and the sample response is measured in the cross-polarized component. While this lowers $B_1$ by a factor of $\sqrt{2}$, and reduces the response (echo) amplitude by the same factor, it provides a 20-30 dB isolation from the pulses, and brings the reflected power below the damage threshold of the mixer-detector.

In cw mode the reflected co-polarized component is used to generate a low phase-noise local oscillator signal for the superheterodyne detection scheme. For pulsed applications, however, there is only appreciable power in the co-polarized component at the time of the pulses, not at the time of the sample response, making a similar superheterodyne detection scheme not possible.

A simple solution is to use a non-phase-sensitive heterodyne detection scheme. The principal detector is a fundamental single port mixer-detector with a local oscillator signal generated by a multiplied Gunn diode at a frequency 5-7 GHz below the operating frequency. The intermediate frequency (IF) signal from this detector is then amplified and rectified with a simple tunnel diode detector (Advanced Control Components).

A clear disadvantage is that the signal is then proportional to the power rather than the amplitude of the response signal, which limits the signal-to-noise for small amplitude signals. Phase-sensitive pulsed detection is preferable, and this mode of operation is shown schematically in Figure 1.

The primary source (S2) consists of a multiplier chain following a phase-locked oscillator in the 12-15 GHz range. The primary detector (M1) is a single-ended Schottky fundamental mixer, which is biased by a DC voltage. A cw Gunn diode (S1) serves as the local oscillator, operating at a frequency of 5 GHz below or above the frequency of the pulsed source S2.

The resulting IF frequency of 5 GHz is then mixed with a 5 GHz source generated by a synthesizer. A millimeter wave single-ended Schottky diode mixer (M1) is used as a detector and two balanced mixers (M2 and M3) are used to separate the signal into two components ($V_s'$ and $V_s''$) with a 90 degree phase difference. Although all frequencies are derived from the same 10 MHz quartz oscillator, the resulting phase noise at these high frequencies is severe. It is therefore currently not possible to measure separately the real (dispersive) and imaginary (absorptive) parts of the susceptibility, as is possible in cw operation. Instead, it is necessary to calculate the phase-insensitive amplitude of the response (echo) by adding the two components in quadrature $V_s(t) = \sqrt{(V_s'(t))^2 + (V_s''(t))^2}$ for each shot. The resulting $V_s(t)$ can then be averaged. In what follows this method will be referred to as quadratured superheterodyne detection. It should be noted that its implementation in existing spectrometers could be useful, as the phase noise in general increases with the distance between pulses, and can lead to significant artifacts in systems with long $T_2$ times. All phase information is lost, but for this spectrometer it is not crucial as there is currently no phase-control of the pulses.

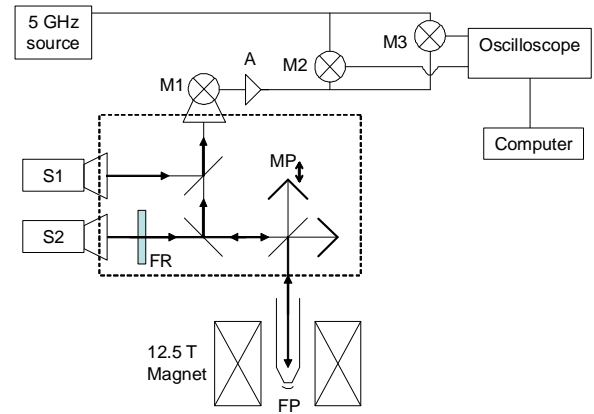

FIG 1. Schematic of the multi-high frequency pulsed EPR spectrometer set up for phase-coherent detection as described in the text. For diode detection the output of the amplifier A is fed into the oscilloscope via a diode. A Faraday rotator (FR) is used as an isolator to protect the pulsed millimeter wave source (S2) from large amounts of unwanted millimeter wave power reflecting back from the Fabry-Pérot (FP) resonator. The Martin Puplett (MP) interferometer is used to control the polarization of the millimeter waves entering the cryostat. A cw millimeter wave source (S1) is used as a local oscillator. M1 is a single-ended millimeter wave Schottky mixer. M2 and M3 are balanced mixers.

Three sources of GHz radiation were purchased from VDI[22]. One provides 240 and 120 GHz, another produces 334, 221 and 112 GHz, and the third creates 336 GHz only. Around 3 mW (30 mW, 100 mW) of power is available at 334 GHz (221 GHz, 112 GHz).

Radio frequency (RF) radiation for ENDOR experiments can be introduced to the sample space via a semi-rigid co-axial cable. It is also possible to introduce light via fibre-optic cables or a glass rod. Eight electrical wires access the sample space.

To upgrade to pulsed operation a larger $B_1$ field is needed so a new Fabry-Pérot resonator was built. A schematic is shown in Figure 2; it is similar to that mentioned in references [1] and [24], with the position of the lower mirror being tunable by rotating a rod that reaches out of the sealed cryostat. This makes it possible to perform experiments at temperatures from 1.8 - 330 K. A piezo-electric transducer is used for fine-tuning of the mirror position by up to 100 μm. The mirror is a fused silica planoconcave lens coated with gold on its spherical side.

The stationary semi-transparent mirror should transmit ~0.2-1% of the incident millimeter wave radiation. For this purpose, meshes were made by photolithography on fused silica substrates to have various linewidths and spacings, appropriate for the full range of frequencies used. These acted as the stationary mirror in the resonator. 15 nm of chromium was evaporated onto 160 µm fused silica cover slips, and then 350 nm of gold was evaporated onto the chromium layer. For 334 (220, 110) GHz, square grids were used with line thickness (and line spacing) of 2.5 (64, 100) µm.

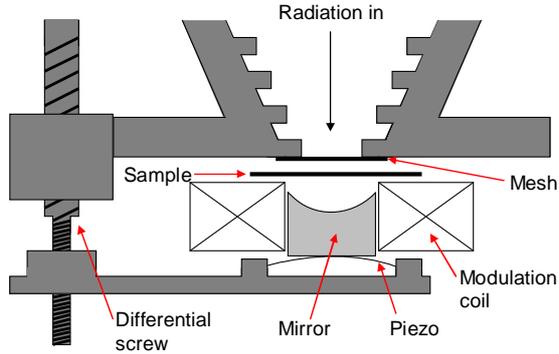

**FIG 2. Schematic of semi-confocal Fabry-Pérot resonator. The differential screw and the piezo-electric transducer (piezo) are used to tune the position of the lower mirror.**

Room temperature frequency sweeps suggest that the quality factor is Q ~ 500 with between five and eight half-wavelengths in the resonator.

## III. RESULTS

Several samples were studied to demonstrate the capabilities of this spectrometer. The typical duration of a π/2-pulse was 600 (300, 100) ns for 334 (221, 110) GHz. The data were collected in heterodyne mode for most of the results shown here, except for the results described in Figure 5 in which the quadratured super-heterodyne detection was used.

The sensitivity[9,14] was measured with a 0.1% (by mass) sample of the organic radical TEMPOL ($C_9H_{18}NO_2$) in a polystyrene film. It has been suggested that TEMPOL may have a role in treating cancer[25]. The sample contained around $4 \times 10^{14}$ spins. Figure 3 shows a single shot echo obtained at 5 K with 239.04 GHz radiation.

The pulses were chosen to maximize the echo and were both about 650 ns 2π/3 pulses, covering a microwave field of ~ 0.04 mT. The cw linewidth of the spectrum was ~38 mT so about 0.1% of the spins participate in the echo. The signal-to-noise ratio (SNR) of a single shot (Figure 3) is 20. Finally, the pulse separation was $\tau$ = 1.5 µs, which is similar to the transverse relaxation time, measured as $T_2$ = 1.49 µs. The

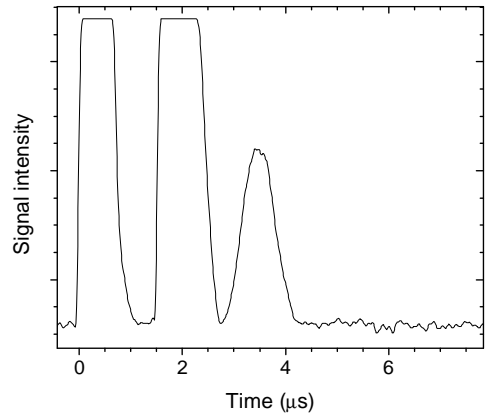

**FIG 3. Single shot echo from TEMPOL sample at 5 K with 239.04 GHz radiation.**

absolute pulsed sensitivity at 240 GHz for a single shot is then calculated as:

$$\text{sensitivity} = \frac{Nf}{SNR} e^{-2\tau/T_2} = 4 \times 10^9 \text{ spins}, \quad (1)$$

where the number of spins is $N$ and the fraction that participate is $f$. At 334 GHz the sensitivity was measured as $2 \times 10^{10}$ spins.

Figure 4 presents the dependence of the Hahn echo decay on magnetic field for the same sample. The g-anisotropy is clearly seen and the $T_2$ time depends on the position in the resonance, as found by Hertel *et al.* at 180 GHz[14].

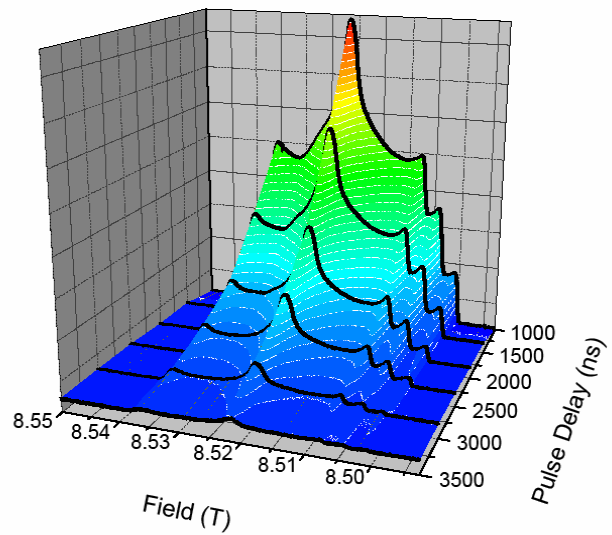

**FIG 4. 2D plot showing field-dependence of Hahn echo decay in TEMPOL at 15 K with 239.04 GHz pulses.**

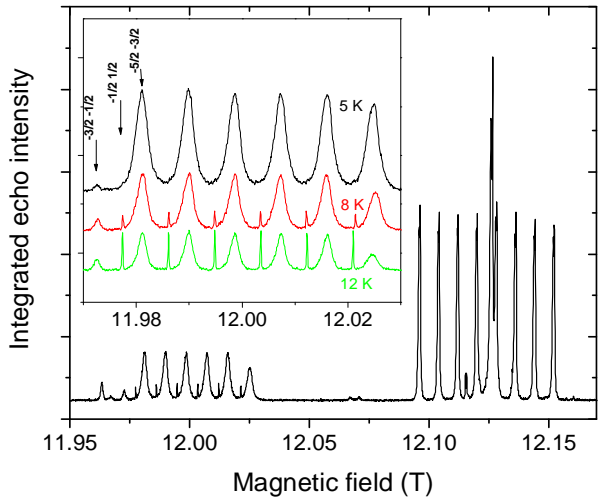

**FIG 5.** Electron Spin Echo (ESE) detected EPR spectrum of a $5.4 \times 10^{-3}$ mm$^3$ (20 µg) MgO single crystal at 336.00 GHz and 8 K. This single scan used 20 shots per point. The first (200 ns) pulse was separated from the second (300 ns) by 1000 ns. The main features are due to Mn$^{2+}$ (electron spin S=5/2, nuclear spin I=5/2) and V$^{2+}$ (S=3/2, I=7/2) impurities. The inset shows the Mn$^{2+}$ spectrum at 5, 8, and 12 K, showing the large electron spin polarization. Only three of the five electron spin transitions are visible, as indicated for the low-field hyperfine components.

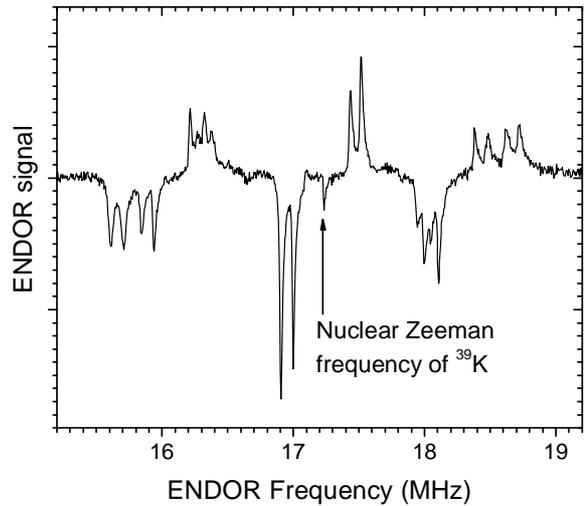

**FIG 6.** Mims ENDOR from a single crystal of Cr:K$_3$NbO$_8$. The signal is due to the 0.1% of Cr$^{5+}$. The temperature was 5 K, the microwave frequency was 240 GHz and the first two pulses were separated by 600 ns.

Figure 5 shows the ESE detected spectrum at 336 GHz of impurities in a 'nominally pure' MgO single crystal with quadratured super-heterodyne detection at 8 K. The small single crystal (0.5×0.18×0.06 mm) was attached with a tiny amount of vacuum grease in the center of the curved mirror of the Fabry-Pérot resonator. At these frequencies the signals from the different impurities are well separated. The main features are the six Mn$^{2+}$ hyperfine components in the 11.97-12.03 T range and the eight V$^{2+}$ components in the 12.09-12.16 T range, with other signals likely due to Fe$^{3+}$ and Cr$^{3+}$. The large electron spin polarization at these temperatures is illustrated in the inset for the d$^5$ Mn$^{2+}$ system. At 5 K the $m_s = -5/2 \leftrightarrow m_s = -3/2$ transition dominates the spectrum with a very small contribution from the $m_s = -3/2 \leftrightarrow m_s = -1/2$ transition. At higher temperatures the narrow $m_s = -1/2 \leftrightarrow m_s = +1/2$ transition is also clear.

Figure 6 is a Mims ENDOR spectrum of potassium nuclei in a single crystal of the peroxy compound Cr:K$_3$NbO$_8$ that was grown as described previously[26]. The concentration of Cr$^{5+}$ (S=1/2) was 0.1%. The unmatched ENDOR 'coil' consisted of a thin copper wire held 1 wavelength (1.25mm) from the mesh, and perpendicular to the polarization direction of the incident millimeter waves. To keep it straight it is supported by threading it through a quartz capillary. The single crystal was attached to the outside of this capillary. Whereas cw ENDOR could not be detected and Davies ENDOR was very weak, a Mims ENDOR sequence provided the clear spectrum shown in Figure 6 in less than thirty minutes. The maximum ENDOR effect was of the order of 50%.

The pulse sequence consisted of two 240 GHz π/2 pulses of length 240 ns separated by tau=600 ns followed by a 150 µs RF pulse, and finally a 240 GHz π/2 pulse to generate the stimulated echo. Around 10 W of RF power was used.

In this spectrum eight equivalent I=3/2 $^{39}$K nuclei are made inequivalent by the applied magnetic field, $B_0$. The quadrupolar coupling of these nuclei dominates the observed splitting. At X-band frequencies all these resonances would be in the 0-3 MHz range where the signal-to-noise is intrinsically poor[15]. However, at 240 GHz the nuclear Zeeman interaction is dominant and brings the ENDOR signals out to a much more accessible frequency range. The interpretation is also simplified by the fact that the EPR level splitting is larger than the thermal energy ($g\mu_B B_0 > kT$), and that the sign of the ENDOR signal changes across the spectrum due to nuclear polarization[18,19,27]. This also leads to a straightforward determination of the sign of the hyperfine splitting. It should be noted that the electronic $T_1$ time, which is ~1 second at helium temperatures[28] at X-band frequencies is considerably shorter at 240 GHz, allowing for a shorter shot repetition time. A detailed study of the high-field ENDOR and relaxation of $Cr:K_3NbO_8$ will be published separately[29].

In conclusion, we have described a quasi-optical multi-frequency (114-336 GHz) pulsed EPR and ENDOR spectrometer. This is particularly suited for studying systems with large zero-field splittings and resolving small *g*-factor anisotropies. The field-dependence of relaxation times can be studied, including the regime where electronic spins are thermally polarized to over 99%.

## ACKNOWLEDGEMENTS

This research was supported by an IHRP grant from the NHMFL, a PEG grant from the Florida State University Research Foundation, NSF grant DMR-0520481 and the EPSRC by grants GR/S23506 and EP/D049717. The National High Magnetic Field Laboratory is supported by National Science Foundation Cooperative Agreement No. DMR-0084173, by the State of Florida, and by the DOE. We would like to thank M. Pati and N.S. Dalal of the Department of Chemistry and Biochemistry at Florida State University for the $Cr^{5+}:K_3NbO_8$ sample.

\* Current address: g.morley@ucl.ac.uk, London Centre for Nanotechnology and Department of Physics and Astronomy, 17-19 Gordon Street, London WC1H 0AH, UK.